\documentclass[12pt]{article}
\usepackage{times}
\usepackage{geometry}
\usepackage[utf8]{inputenc}
\usepackage{enumitem,amssymb}
\usepackage{amsmath}
\usepackage{graphicx}
\usepackage{ragged2e}
\usepackage[dvipsnames]{xcolor}
 \usepackage{hyperref}
\newlist{thematic}{itemize}{8}
\setlist[thematic]{label=$\square$}
\usepackage{pifont}
\newcommand{\cmark}{\ding{51}}%

\newcommand{\checkbox}{\rlap{$\square$}{\raisebox{2pt}{\large\hspace{1pt}\cmark}}\hspace{-3pt}}

\newcommand{\etal}{\textit{et~al.}}

\begin{document}
\raggedright

\Large
{Astro2020 APC White Paper}\\[18pt]

\huge
Entering into the Wide Field Adaptive Optics Era on Maunakea\\[22pt] 

\normalsize

\noindent \textbf{Type of Activity:}\\[2pt] \hspace{0.1cm}
\checkbox\ Ground Based Project \hspace{0.3cm} 
$\square$ Space Based Project \hspace{0.35cm} 
\checkbox\ Technological Development Activity \\ \hspace{0.1cm}
\checkbox\ Infrastructure Activity  \hspace{0.25cm}
$\square$ State of Profession Consideration \hspace{0.5cm}
$\square$ Other \linebreak
  
\textbf{Principal Author:}

Name:	Gaetano Sivo  \linebreak
Institution:  Gemini Observatory  \linebreak
Email: gsivo@gemini.edu  \linebreak
Phone:  +56 51 2205 642  \medskip
 
\justify\textbf{Co-authors:} John Blakeslee (Gemini), Jennifer Lotz (Gemini), Henry Roe
(Gemini), Morten Andersen (Gemini), Julia Scharw\"achter (Gemini), David Palmer (Gemini),
Scot Kleinman (Gemini), Andy Adamson (Gemini), Paul Hirst (Gemini), Eduardo Marin (Gemini),
Laure Catala (Gemini), Marcos van Dam (Flat Wavefronts), Stephen Goodsell (Gemini),
Natalie Provost (Gemini), Ruben Diaz (Gemini), Inger Jorgensen (Gemini), Hwihyun Kim
(Gemini), Marie Lemoine-Busserole (Gemini), Celia Blain (Gemini), Mark Chun (IfA -
U.\ Hawai'i), Mark Ammons (LLNL), Julian Christou (LBTO), Charlotte Bond (IfA -
U.\ Hawai'i), Suresh Sivanandam (U.\ Toronto), Paolo Turri (UC, Berkeley), Peter
Wizinowich (Keck), Carlos Correia (Keck), Benoit Neichel (LAM), {Jean-Pierre} V\'eran
(NRC-Herzberg), Simone Esposito (INA, OAA), Masen Lamb (U.\ Toronto), Thierry Fusco
(LAM), Fran\c cois Rigaut (ANU), Eric~Steinbring (NRC-Herzberg) \par\bigskip

\justify\textbf{Abstract:}
As part of the National Science Foundation funded ``Gemini in the Era of
Multi-Messenger Astronomy'' (GEMMA) program, Gemini Observatory is developing GNAO, a wide-field
adaptive optics (AO) facility for Gemini-North on Maunakea. GNAO will
provide the user community with a queue-operated Multi-Conjugate AO (MCAO) system,
enabling a wide range of innovative solar system, Galactic, and extragalactic science with
a particular focus on synergies with JWST in the area of time-domain astronomy.
The GNAO effort builds on institutional
investment and experience with the more limited block-scheduled Gemini
Multi-Conjugate System (GeMS), commissioned at Gemini South in 2013. The project 
involves close partnerships with the community through the recently
established Gemini AO Working Group and the GNAO Science Team, as well as external
instrument teams.  The modular design of GNAO will enable a planned upgrade to a Ground
Layer AO (GLAO) mode when combined with an Adaptive Secondary Mirror (ASM). By enhancing
the natural seeing by an expected factor of two, GLAO will vastly improve Gemini North's
observing efficiency for seeing-limited instruments and strengthen its survey capabilities
for multi-messenger astronomy.  \par\bigskip
\parskip 2pt
\section{Introduction}\label{sec:intro}

Over the past two decades, major Observatories in both the Northern and Southern
Hemispheres have developed Adaptive Optics (AO) instrumentation in order to achieve
near-diffraction limited observations for members of the astronomical community whose
science depends on high spatial resolution. The coming generation of extremely
large telescopes (ELT) such as the Thirty Meter Telescope (TMT), Giant Magellan Telescope,
and European Extremely Large Telescope are all designed with AO intrinsically embedded
into their operations. Thus, AO will be a central part of their standard observing modes
and has been a strong factor in optimizing their designs. However, these are not likely to
be available to the US community before the end of the coming decade.

In parallel, the US community will also have access to two revolutionary new
optical-infrared facilities: the Large Synoptic Survey Telescope (LSST) and the James Webb
Space Telescope (JWST). These two telescopes will enable unprecedented science at high
resolution (in the case of JWST) and in the time domain (with LSST). However, neither of
these important facilities will be able to provide high-cadence long-duration monitoring
of variable phenomena at high angular resolution. Achieving this requires
a wide-field AO system fully integrated into the operations of a major observatory  so
that the capability is available on any night with suitable conditions.
Wide Field Adaptive Optics systems, such as Multi-Conjugate AO (MCAO: Beckers 1988, Rigaut
$\&$ Neichel 2018), Laser Tomographic AO (LTAO: Tallon $\&$ Foy 1990), Multi-Object AO
(MOAO: Gavel 2004), and Ground Layer AO (GLAO: Rigaut 2001) represent the path forward.

Gemini Observatory's twin 8.1-meter telescopes, located on prime observing sites
in Hawaii and Chile, provide forefront access to astronomical targets over the entire sky. 
Gemini South (GS) is situated on Cerro Pachon at an altitude of 2722\,m and latitude
$-$30.2~deg, while Gemini North (GN) is located on Maunakea at 4213\,m and $+$19.8~deg.
%
Gemini provides open-access 8m-class telescopes for the US astronomical
community; it is essential to maintain outstanding instrumentation and operational
performance to serve the diverse needs of this community.
To meet this need, the NSF has awarded the multi-million dollar Gemini in the Era of
Multi-Messenger Astronomy (GEMMA) program to advance Gemini's leadership in the areas of
multi-messenger, time-domain astronomy and high angular resolution science.
GEMMA funds key aspects of Gemini's
\textcolor{blue}{\href{https://www.gemini.edu/files/general-announcements/gemini-strategic-plan.pdf}{Strategic
 Scientific Plan for the 2020s}}
and is designed to maximize the synergies with major new facilities
such as JWST and LSST that will revolutionalize high-resolution and time-domain studies in
the coming decade, as well as multi-messenger facilities such as LIGO/Virgo.

\begin{figure}[!ht]
\centering    
\includegraphics[width=3.85in]{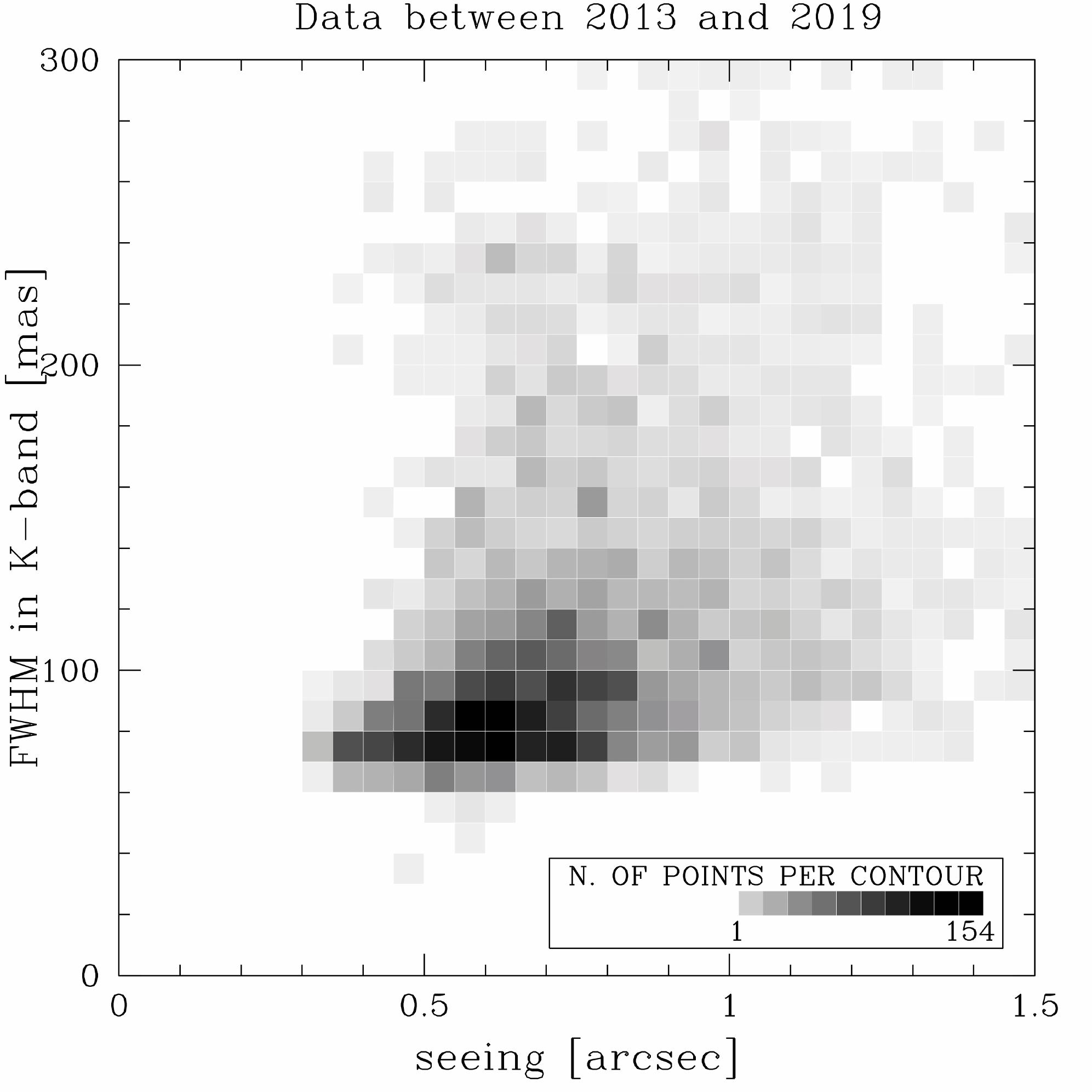}\vspace{-2mm}
\caption{Probability density FWHM values measured on GeMS/GSAOI science exposure in the
  $K$-band since GeMS commissioning, plotted as a function of the estimated $V$-band seeing.}
\label{fig:FWHM_K}
\end{figure}

The Gemini telescopes were designed for optimal image quality and are outfitted with facility AO
systems at each site.
At GS, the multi-conjugate GeMS AO system (Rigaut \etal\ 2014) uses a unique constellation
of five laser guide stars (LGSs), and up to three natural guide stars (NGSs).  Coupled
with the Gemini South AO Imager (GSAOI), GeMS takes advantage of Gemini's design strength to
deliver nearly diffraction-limited $K$-band images with typically 0.085$''$ resolution over a
1.4$'$ field of view (see Fig.~\ref{fig:FWHM_K}).
In contrast, although Maunakea is a superior site for AO performance, GN lacks any wide-field
AO capability. Its narrow-field single-conjugate ALTAIR system, which was among the first
facility LGS AO systems in routine operation,
falls well short of fully exploiting the site’s outstanding characteristics. 
Therefore, as part of the NSF-funded GEMMA program, Gemini is developing a new wide-field 
MCAO facility for GN, known simply as GNAO, with a path towards an even wider field GLAO system.
We discuss the key science drivers of the GNAO project in the following section.

%
%

\medskip

\section{Key Science Goals and Objectives}



Although the development of the new multi-conjugate GNAO system will be enlightened by
experience from GeMS, the system will be far more advanced.
Operating from Maunakea with an improved design, GNAO will achieve superior performance
including diffraction-limited imaging in the $K$ band over a $\sim\,$1.5$'$
field. However, the most dramatic difference in requirements pertains to the system
reliablility, ease of operations, and nightly availability.

Unlike Gemini's other facility instruments, which are operated from the base facility by a
crew consisting of a telescope operator and a queue observer, the GeMS/GSAOI system is currently operated
from the summit and requires a team of four; thus, it is typically scheduled for only two
or three one-week runs per semester. In contrast, GNAO is designed to be queue operable
by the standard two-person crew without additional support. Thus, it will be available for
observations on any night that the telescope is open, and 
this makes a world of difference in the variety of science that it enables.
In addition to the high-resolution studies of stellar populations, supernova environments,
proper motions, and galactic archaeology that have been done with GeMS, the nightly  queue
scheduling capability of  GNAO opens the time domain to detailed high-resolution
investigations. 

Examples of such high-resolution, time-domain science include monitoring long-period
variables in galaxies as distant as the Virgo cluster to constrain lifetimes on the thermally
pulsating asymptotic giant branch in diverse stellar populations, searches for intermediate-mass
black holes from the internal motions of dense stellar systems, rest-frame optical studies
of galaxy interactions at the peak of cosmic star formation activity near $z\sim2$, and
exploration of the earliest stages of galaxy formation via narrow-band imaging of
Lyman-alpha emitters at redshifts where the emission falls between the atmospheric OH
lines.

These science topics will also be investigated by JWST.
However, as discussed in more detail by the Astro2020 science white paper by Blakeslee
\etal\ (2019), because of JWST's limited range of pointing angles, very few Target of
Opportunity (ToO) observations, and extremely high proposal pressure, there will be very strong 
synergy between Gemini/GNAO and  JWST.  GNAO will be the only system
able to study and monitor high-priority northern targets with a similar spatial resolution
and field of view as the Near InfraRed Camera (NIRCAM) on JWST during months when
those targets are not observable by JWST itself. This will be especially important for
time-domain science including lensed supernovae (e.g., Kelly
\textit{et al.} 2015), for which additional lensed images are predicted to appear
depending on the lens model when the target may be unobservable by JWST and when Hubble
itself may no longer be operational. 
GNAO will also enable high-resolution investigation of high-priority targets
identified by LSST visible from both hemispheres, in some cases while they are being observed
spectroscopically from GS. Combined with GeMS, the GNAO system will allow Gemini
to provide to its community unique wide-field AO over the entire sky.
In addition, GNAO will support of the next generation of LGS AO-assisted instruments bound for GN,
including the ambitious Gemini InfraRed Multi-Object Spectrograph (GIRMOS; Sivanandam
\etal\ 2018), an externally funded multi-IFU spectrograph being built for Gemini by a
Canadian-led instrument team. 

The long-term plan for GN calls for the deployment of an Adaptive Secondary Mirror (ASM)
in order to increase the efficiency of all observations conducted at GN, with and without
GNAO.  The number of necessary optical surfaces in an AO system decreases with the 
addition of an ASM,
limiting the loss from thermal emissivity and improving the overall throughput. An ASM
also facilitates correction over a very wide field ($> 8'$) using a GLAO system.
Essentially, a GLAO system provides ``super-seeing'' observations,
with 2 to 3 times better resolution compared to natural seeing, or the equivalent in terms
of signal-to-noise for point sources of doubling the primary mirror collecting area.
Indeed, the GLAO correction greatly enhances the efficiency of almost all observations done
on 8m-class telescopes and enables a range of science that can only
be achieved with high spatial resolution.  A powerful wide-field GLAO system, enabled by a
state-of-the-art ASM on the 8m GN telescope, 
coupled to the GNAO system for diffraction-limited science over a smaller 1$'$--2$'$ field,
will be an extremely powerful and efficient capability for  the US  community
in advance of the ELT era and would create
synergies with those facilities when they come online.

While the GEMMA-funded baseline GNAO design does not include any GLAO specific components such
as an ASM or very high order wavefront sensor embedded into the Acquisition and
Guiding unit of the telescope, the design is being optimized in order to accommodate such
upgrades once the GLAO program can become a reality.  During the Conceptual Design Stage of
the project, the team will engage with our community to evolve the science cases prepared
and gathered during a GNAO workshop held several years ago to identify further science cases.  Based
on past experience, we have formed an initial core science team comprised of members
internal and external to Gemini that will help us further refine a set of research
objectives resulting in a science cases document at the end
of the Conceptual Design Stage.

\section{Technical Overview: from GNAO to its upgrades path}
The new GNAO system is composed of three main subsystems: 
\begin{itemize}
    \item an Adaptive Optics bench;
    \item a Laser Guide Star facility;
    \item a Real Time Computer.
\end{itemize}
We describe each of these in turn.

\subsection{GNAO: Adaptive Optics System}
\begin{figure}[!ht]
    \centering
    \includegraphics[width=5in]{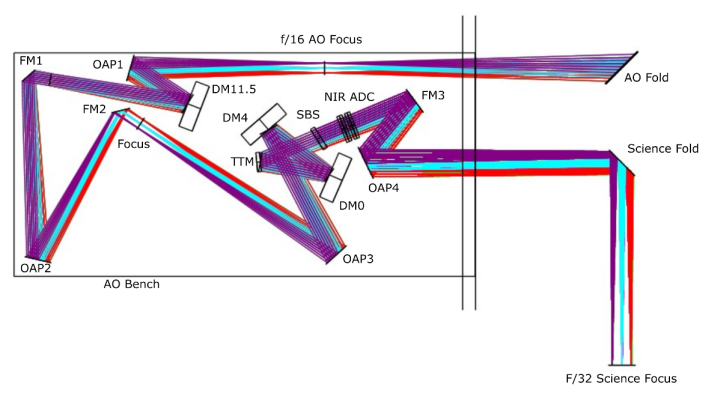}
    \caption{Conceptual optical design for the AOS Science Path, showing 3 DMs. }
    \label{fig:AOS}
\end{figure}

The Adaptive Optics Subsystem (AOS), sometimes also referred to as the AO Bench (AOB), will measure wavefront aberrations introduced mostly by the Earth's atmosphere and apply corrections for these aberrations.  Since the AOS needs to handle multiple LGSs and NGSs, the design and implementation of the AOS is very challenging, particularly within Gemini’s volume and mass constraints.  
In order to adequately correct the atmosphere over a $\sim\,$2$'$ wide field, as is the goal for MCAO, multiple deformable mirrors (DMs) are required.  As shown in Figure \ref{fig:AOS}, three DM conjugation locations are employed, although one will be a flat mirror in anticipation of the planned ASM.
Since most of the turbulence is ground layer (especially on an excellent  site like Maunakea), it is critical to have one DM conjugated to 0~km. When Gemini is able to deploy an ASM into its structure, the ASM will play the role of the DM conjugated to the ground, and thus we can re-use the DM0 inside the AOS to conjugate to a higher altitude (here the mid-layer one is at 4~km). This can  be achieved with a suitable AOS design that we have developed.

\subsection{GNAO: Laser Guide Star facility}
The LGSF uses lasers to cause sodium in the upper atmosphere to fluoresce and, thereby, forms `stars' that the AO subsystem (AOS) can use to correct atmospheric turbulence (instead of natural guide-stars that are only sparsely available with adequate brightness).  For this MCAO system, we envision to create an artificial constellation of five laser guide stars on-sky coming from three Toptica lasers (Enderlein \textit{et al.} 2014). Two laser beams will each be split in two and launched from the side of the telescope, and one laser launched from the center as it is done currently on the ALTAIR system. 
\begin{figure}[t]
    \centering
    \includegraphics[width=5.95in]{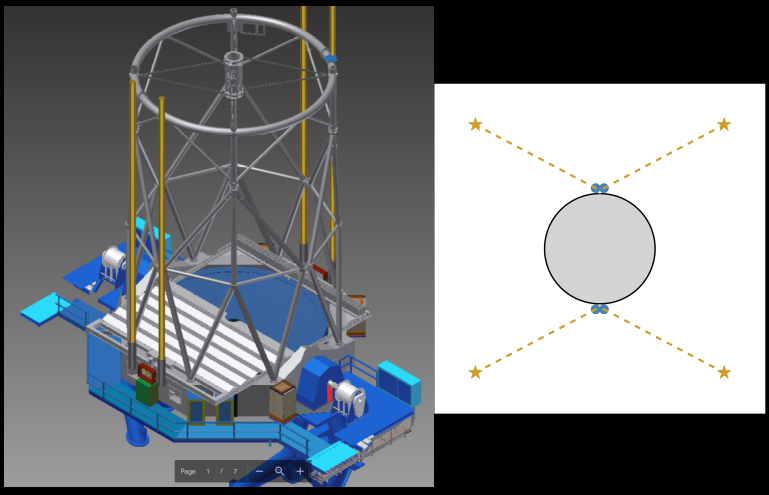}\vspace{-4pt}
    \caption{Mechanical view of the laser guide star facility}
    \label{fig:LGSF}
\end{figure}
The LGS constellation is 45$''$ radius wide. But the main requirement in the design is to be able to expand the constellation radius up to 7$''$ in order to enable very wide field constellation that could be used for a very wide field ground layer AO system in the future. 

\subsection{GNAO: Real Time Computer}
The Real-Time Computer (RTC) reads the wavefront sensor (WFS) data, calculates corrections, and outputs these corrections to the deformable mirrors (DMs) and tip/tilt (TT) mirror.  This all needs to be done at 500 frames per second (fps) or faster.  Given the required frame rate, the number of WFSs, and the number of DMs, a huge amount of input/output and large numbers of calculations need to be done extremely quickly.  In addition, other functions need to be performed, including interfacing with the outside world; providing streams of data for diagnostic, operations, and analysis purposes; and updating wavefront reconstruction matrices (that are used in calculating corrections). Thanks to the GEMMA funding, a powerful and modular RTC platform will be developed for GNAO and will be available for the future GLAO system. 

\subsection{GNAO limitation and path forward}
GNAO is designed to be an MCAO facility delivering  diffraction limit performance over a wide but still limited field of view ($\sim 2'$). Maunakea is an excellent site for astronomical observations, and in particular its superb seeing conditions and turbulence profile (Tokovinin \textit{et al.} 2005) enable another era that would increase significantly an 8-m telescope efficiency: Ground Layer Adaptive Optics (GLAO). This was actually already envisioned back in 2003 when Gemini released a call for proposal for a feasibility study for a GLAO study. The conclusion at the time was that a Ground Layer Adaptive Optics (GLAO) system for Gemini would offer very significant improvements in image quality and observing efficiency. A GLAO system would enable new science cases for Gemini such as: \vspace{-3pt}
\begin{itemize}
    \item Efficient surveys of large areas for first light objects; deployable IFUs (d-IFUs)
    and the improved GLAO image quality will produce a significant multiplexing advantage for surveys of velocity fields which will shed light on dark matter on galactic scales; finally the improved image quality and uniform point spread function (PSF) produced by GLAO will facilitate proper motion studies within the Local Group. Even while these specific Aspen science cases are addressed, an important feature is that every non-diffraction limited Gemini science proposal will benefit from a GLAO facility. 
    \par\vspace{-2pt}
\item Image quality statistics will be drastically altered by GLAO; image quality conditions which occur only $20\%$ of the time currently, will occur $60-80\%$ of the time when GLAO is operational. The GLAO-improved seeing statistics will ensure that top ranked proposals which require good image quality will be successfully observed and, in general, will ease scheduling constraints and improve the operational efficiency of the observatory. Improved image quality also translates into shorter exposure times and an increase in the number of programs executed. Gemini should realize a net $50\%$ improvement in overall efficiency. 
\par\vspace{-2pt}
\item The large GLAO-corrected field of view and a potential d-IFU spectrograph will make Gemini a more efficient survey telescope, while improving the efficiency and performance of every seeing limited instrument at all available wavelengths. 
\end{itemize}
This all become possible if Gemini moves forward with the deployment of an ASM.
At the time of Gemini's earlier GLAO study, the feasibility of this was still a major concern as the technology was not proven yet. Today it is a different story. Several major observatoris now own and operate ASMs (MMT, Magellan, both sides of the Large Binocular Telescope, and one of the UT units at the European Southern Observatory). 

\begin{figure}
\begin{tabular}{cc}
\includegraphics[width=3.2in]{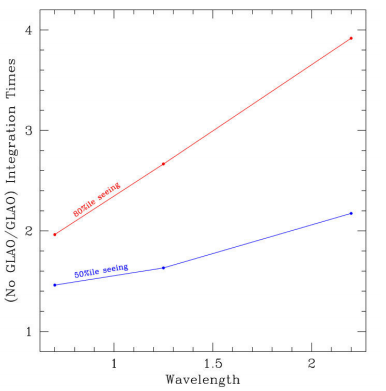} &\includegraphics[width=2.9in]{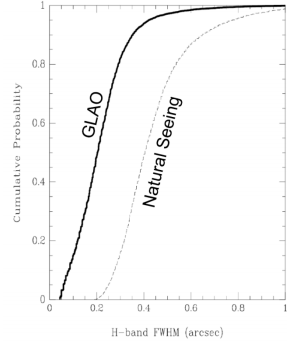}
\label{tab:GLAO2} 
   \end{tabular}
\caption{Left: the reduction in integration times from 0.7 to 2.5$\mu$m with a GLAO system
  at Gemini to achieve the equivalent point source sensitivity under $50^{\text{th}}$ and
  $80^{\text{th}}$ percentile seeing.
  Right: the predicted decrease in PSF size at $1.6\mu$m for a GLAO system at Gemini-N is shown. To first order, a GLAO system would shrink the PSF by a factor of at least 2 under all natural seeing conditions.}\label{fig:glao}
  \end{figure}

A GLAO system at GN driven by a state-of-the-art ASM would revolutionize the observatory's operating efficiency. Indeed, on the left side of the Figure~\ref{fig:glao}, we see the typical boost in point source sensitivity that is expected with a GLAO system. The right side of the figure illustrates the cumulative probability of seeing conditions with and without GLAO.

\section{Technology drivers}
The technology involved in the MCAO development is well proven by  GeMS.
While adaptive secondary mirrors are less well established, the technology became a reality in term of reliability, ease of use, and complexity several years ago. Indeed the successful development, integration and commissioning of the unit developed for ESO at the VLT has changed the view of such systems (Gallieni \textit{et al.} 2014). The coming generation of ELTs are all designed with AO intrinsically embedded into their operations. Thus, AO will be a central part of their standard observing modes and has been a strong factor in optimizing their designs. Deformable mirrors are directly embedded into the telescopes designs. There is an opportunity now to enable Gemini on this path as well. 

Given the complex and highly specialized engineering required to build an ASM, this
subsystem will be procured via competitive bid world-wide. Particularly important at this
point will be packaging the entire system, including all off-telescope calibration and
alignment equipment needed for the ASM. The space demands of the system will be meshed
with space requirements in the complex and densely populated M2 region of the
telescope. Trades will be made between the nominal heat budget of the system, loads on the
cooling capacity at M2, space required of the central laser launch system, methods to
retrofit a new hexapod mount of the M2 assembly into the existing structure (assuming this
mounting scheme is adopted), the optimal location of digital and analog electronics, face
sheet thickness versus actuator density, etc. An end-to-end error budget will be kept at
Gemini so that these design trades are reflected in a single methodically managed format
to which all subsystem developers will contribute.

The ASM on Gemini would become intrinsically a standard operation part of its structure and is envisioned to work on daily basis for a lifetime of more than 20 years.  

\section{Organizations, Partnerships and current Status}

\begin{figure}[!ht]
  \centering
  \includegraphics[width=6in]{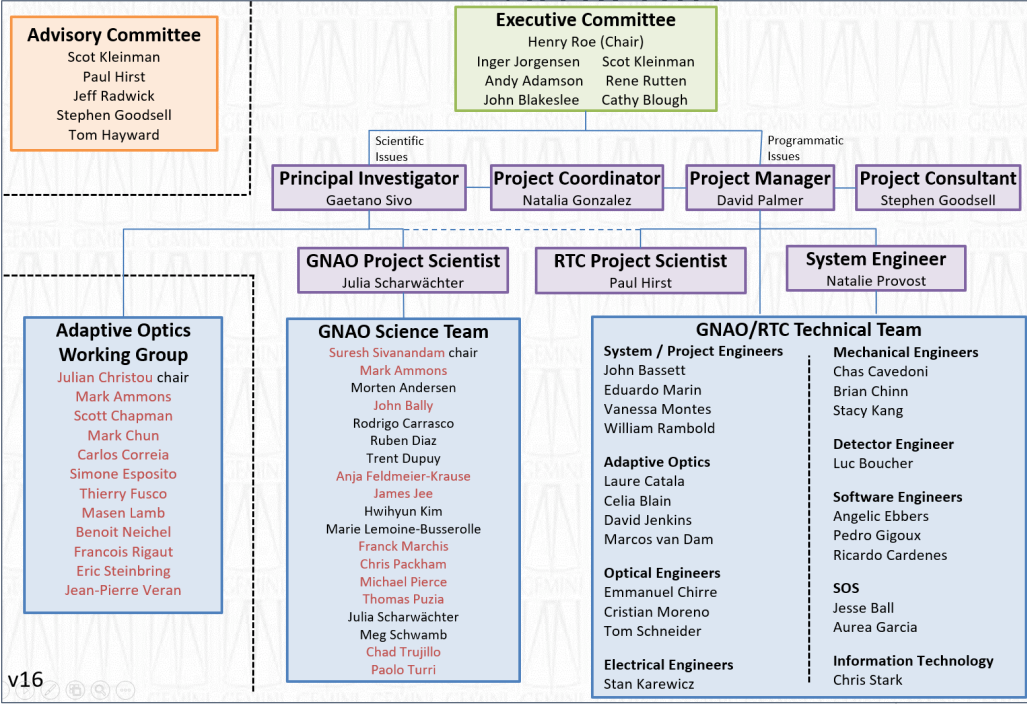}\vspace{-2pt}
 \caption{Organizational chart of the GNAO team. Black represent staff from Gemini
   Observatory; red represents non-Gemini staff, illustrating the strong community involvement.}
    \label{fig:orgchart}
\end{figure}

The GNAO program is funded and underway.  The envisioned team for the GLAO program will benefit from the legacy of a skilled team put together for the GNAO and RTC development efforts under the GEMMA program. Figure~\ref{fig:orgchart} presents the organization chart of the GNAO team. 

The GNAO Project Manager (PM) and Principal Investigator (PI) share authority and responsibility for leading the project, the PM for programmatic issues and the PI for scientific issues.  The dashed line between the science and technical teams in the org chart indicates that, although responsibility resides on either side of the dashed line, there will be frequent interactions between the two. Although it is not indicated on the org chart, the technical team will report to the System Engineer (SE) for SE related issues and to the PM for programmatic related issues. In practice, this will be a team effort as well.
In this structure, the project is overseen by a Gemini Executive Committee that, in turn, report to the Gemini Director and the NSF.  Specifically, authority flows downward from the Director to the Executive Committee to the PI and PM, then to the rest of the project team. Escalation flows upward along the reverse path. Escalations involving scientific issues involve the Gemini Chief Scientist. Escalations concerning program or portfolio issues  include the Program and Portfolio Managers. Similarly, issues arising from the Portfolio and Program Managers should go from them to the Executive Committee.

In developing GNAO/RTC and later  the GLAO program (if funded), we are seeking to create broad partnerships within the Gemini community.  One initiative is through the re-initiation of the Gemini Adaptive Optics Working Group (AOWG). The AOWG  provides guidance to Gemini and useful information and exploration for working group members. Previous versions of the AOWG have performed complex trade studies and analyses for past AO efforts at Gemini.  We  initially focused the AOWG on helping form top-level requirements for GNAO and the modular RTC platform. The AOWG charter sets member expectations to:
\begin{itemize}
    \item Provide expertise regarding the Observatory’s AO program including technical and design recommendations, community experience, and best practices;
    \item Bring lessons learned from other AO systems so new Gemini systems can build on previous work the observatory has not been directly involved in;
    \item Develop science cases for the observatory AO program that can be used to guide design decisions;
    \item Provide simulations, input parameters, and output results to help develop the AO program.
\end{itemize}
In addition, Gemini has invited several external scientists to join the GNAO Science Team. Finally, the Gemini STAC and Board will also have insight and involvement in these efforts. 

\section{Schedule}
\begin{figure}[!b]
    \centering
    \includegraphics[width=4.5in]{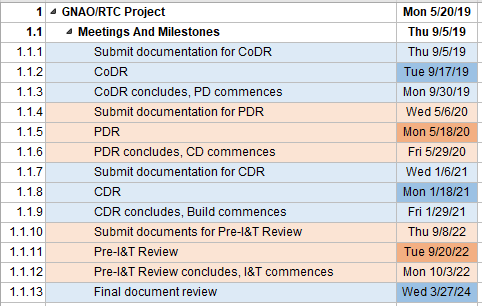}
    \caption{Scheduled reviews of the GNAO project.}
    \label{fig:schedule}
\end{figure}

The figure \ref{fig:schedule} shows the reviews scheduled for the GNAO project. 
The STAC and Gemini Board of Directors endorsed the feasibility study of integrating an ASM at Gemini North. The feasibility study will be funded within the Instrumentation and Development division of Gemini and is expected to be kicked off the third quarter of 2019. If funding is obtained by Gemini to procure an ASM, the schedule would be to start it around the AIT phase of GNAO (2022) in order to start integration on the telescope around 2027. Indeed, it is now known that the lead time for the procurement and build phase of an ASM is several years. We do count for about a year of telescope daytime work for installation and calibration time. Full ASM operation would start in 2028, likely about two years in advance of TMT NFIRAOS commissioning. 

\section{Cost Estimates}

Again, we note that the development of the MCAO system is funded and the GNAO project is progressing towards conceptional design review in late 2019, with PDR scheduled for mid 2020.  Early in the proposal phase of GNAO, the plan was to incorporate an ASM as part of
the system, enabling both MCAO and GLAO corrections.  The plan required redesigning the
Acquisition and Guiding (A$\&G$) unit of Gemini North in order to include the wavefront
sensor of the GLAO mode inside it. Due to resource limitations, in developing the
execution plan, it was decided to descope all GLAO components from the project plan as it
was beyond the amount we could request for the GEMMA program. However, because a GLAO
system enabled by an adaptive secondary remains an essential part of Gemini's strategic
plan, we are designing GNAO such that all relevant parts of the MCAO system (e.g., LGS
facility) will be usable for GLAO as well.  We thus have a good idea of the cost of an
ASM and the redesign work of the A$\&G$. It will require approximately $\$$6M for the
ASM, $\$$5M for the redesign and hardware of the A$\&G$, and about 15 FTEs of effort.  A
total of $\$$14M to commission an equipped GLAO system is then required.

\section{Acknowledgements}
The authors would like the thank the National Science Foundation for the funding of the GEMMA program under the Contract Support Agreement number AST-1839225. 
The Gemini Observatory is operated by the Association of Universities for Research in Astronomy, Inc., under a cooperative agreement with the NSF on behalf of the Gemini partnership: the National Science Foundation (United States), National Research Council (Canada), CONICYT (Chile), Ministerio de Ciencia, Tecnolog\'{i}a e Innovaci\'{o}n Productiva (Argentina), Minist\'{e}rio da Ci\^{e}ncia, Tecnologia e Inova\c{c}\~{a}o (Brazil), and Korea Astronomy and Space Science Institute (Republic of Korea).

\pagebreak
{\par\parskip 8 pt
\noindent\textbf{References}\par

\noindent
Abell, P. A., \textit{et al.} (2009), ``LSST Science Book, Version 2.0,'' arXiv:0912.0201


\noindent
Beckers, J., (1988), ``Increasing the size of the isoplanatic patch with multiconjugate adaptive optics,'' Very Large Telescopes and their Instrumentation, Vol. 2, p 693

\noindent
Blakeslee, J. \etal\ (2019), ``Probing the Time Domain with High Spatial Resolution,''
Astro2020 Science White Paper, BAAS, 51, 529

\noindent
Boyer, C. \textit{et al.} (2019), ``Adaptive Optics Program at TMT,'' AO4ELT6

\noindent
Enderlein, M. \textit{et al.} (2014), ``Series production of next-generation guide-star
lasers at TOPTICA and MPBC,'' SPIE, 914807 

\noindent
Gallieni, D. \textit{et al.} (2013), ``The new VLT-DSM M2 unit: construction and electromechanical testing,'' AO4ELT3

\noindent
Gavel, D., (2004), ``Tomography for multiconjugate adaptive optics systems using laser guide stars,'' SPIE, Vol. 5490, p 1356

\noindent
Ivezi\'c, $\check{\text{Z}}$., \textit{et al.} (2018), ``LSST: from Science Drivers to Reference Design and Anticipated Data Products,'' arXiv:0805.2366v5

\noindent
Kelly, P. \textit{et al.} (2015), ``Detection of a SN near the center of the galaxy cluster field MACS1149 consistent with predictions of a new image of Supernova Refsdal,'' the astronomer telegram

\noindent
Neichel, B., (2014), ``Gemini multi-conjugate adaptive optics system review II: Commissioning, operations and overall performance,'' MNRAS, 440, 1002

\noindent
Rigaut, F., (2001), ``Ground-Conjugate wide field adaptive optics for the ELTs,'' Beyond
Conventional Adaptive Optics 

\noindent
Rigaut, F., (2014), ``Gemini multi-conjugate adaptive optics system review I: Design, trade-offs and integration,'' MNRAS, 437, 2361

\noindent
Rigaut, F. and Neichel, B., (2018), ``Multiconjugate Adaptive Optics for Astronomy,'' ARA$\&$A, 56, 277

\noindent
Saracino, S.  \textit{et al.} (2016), ``Ultra-deep GEMINI Near-infrared Observations of the Bulge Globular Cluster NGC 6624,'' ApJ, 832, 48

\noindent
Sivanandam, S. \etal\ (2018), ``Gemini infrared multi-object spectrograph: instrument overview,'' SPIE, 107021

\noindent
Tallon, M. and Foy, R., (1990), ``Adaptive telescope with laser probe - Isoplanatism and cone effect,'' A$\&$A, 235, 549



\noindent
Tokovinin, A. \textit{et al.} (2005), ``Optical Turbulence Profiles at Mauna Kea Measured by MASS and SCIDAR,'' PASP, 117, 395

\par}

\end{document}